# An Optimized Pattern Recognition Algorithm for Anomaly Detection in IoT Environment

**Nazim Uddin Sheikh[1,2], Hasina Rahman[1] and Hamid Al-Qahtani[1]**
Macquarie University[1], Sydney, Australia
Institute of Engineering & Management[2], Kolkata, India

**Abstract**

With the advent of large-scale heterogeneous search engines comes the problem of unified search control resulting in mismatches that could have otherwise avoided. A mechanism is needed to determine exact patterns in web mining and ubiquitous device searching. In this paper we demonstrate the use of an optimized string searching algorithm to recognize exact patterns from a large database. The underlying principle in designing the algorithm is that each letter that maps to a fixed real values and some arithmetic operations which are applied to compute corresponding pattern and substring values. We have implemented this algorithm in C. We have tested the algorithm using a large dataset. We created our own dataset using DNA sequences. The experimental result shows the number of mismatch occurred in string search from a large database. Furthermore, some of the inherent weaknesses in the use of this algorithm are highlighted.

## 1. Introduction

Extensive usage of string searching algorithms in many branches of science and technology has drawn the attention of engineers and scientists since 1970s. In particular, pattern matching algorithms [1] have been used to solve myriad of problems in many diverse fields such as bioinformatics, network security, and search engine design. Text-editing programs frequently demands to search all possible occurrences of a pattern in the text. Basically, the text is a document or record string that consists of characters being analyzed, and the pattern that is searched for in the text string – is characterized as 'string searching or pattern recognition problem' which is so helpful in text-editing application [2]. Pattern matching algorithms are classified into two categories, namely, exact-pattern matching algorithms and approximate-pattern matching algorithms [3]. The aim of exact pattern matching is to locate all occurrences of an exact pattern in a text and its indices. Some noteworthy exact pattern matching algorithms are Brute-Force algorithm, Knuth-Morris-Pratt algorithm (KMP algorithm) [4], Karp-Rabin algorithm [5], Aho-Corasic algorithm [6], and Boyer-Moore algorithm [7]. This paper is organized as follows. Section II highlights some related work. It addresses some pattern matching algorithms. In section III, we discuss our proposed algorithm and its variants computation formulae. Section IV explains the multifarious applications of our proposed algorithm. We conclude our work and give directions for future research work in Section V.

## 2. Related Work

In this section, we briefly present some of the most popular string recognition algorithms that have been developed in the last few decades. String matching algorithms play an indispensable role in the development of modern technology.

2.1 Boyer-Moore Algorithm

The Boyer-Moore algorithm (B-M algorithm) is considered to be a benchmark in the literature for a long time. The basic principle of the B-M algorithm is follows. It compares the pattern with the text from rightmost character of the pattern to the leftmost character of the pattern. The foundation of this algorithm is based on some shift mappings or functions that maps the shifting index according to characters' match or mismatch occurrence. The B-M algorithm uses two heuristics: good suffix shift mapping and bad character shift mapping. The B-M algorithm is comprised of two phases: preprocessing phase and comparison phase. The time and space complexity in its preprocessing phase is $O(m+\sigma)$, where $\sigma$ is the length of the finite set that compatible with the text and the pattern while the matching phase time complexity is $O(mn)$, worst-case scenario [8] whereas the best-case complexity is $O(n/m)$. The main advantage of this algorithm is the shift rule concept that results some good shift values.

2.2 KMP Algorithm

KMP algorithm [9] is the most efficient and widely used algorithm. The working principle of KMP algorithm is based on longest prefix concept. Its worst-case time complexity is linear [10]. This algorithm does not work well when the size of the alphabet increases that leads to more likely to occur mismatch. The KMP algorithm consists of following two phases. In the preprocessing phase computation, the pattern is processed to determine an array of values derived by using the longest proper prefix which is also a suffix (denoted as LPS). This array of LPS values are utilized to optimize the skips of the pattern in the matching step.

### 2.3 Rabin-Karp Algorithm

The principle behind the Rabin Karp algorithm is based on randomized fingerprint method [11] or hashing technique. To find hash value it uses Horner's rule and hash function. Furthermore, the algorithm applies congruence modulo arithmetic in finding hash value. Though the time complexity is O (mn), it has many applications in different fields such as in checking plagiarism. However, the R-K algorithm exhibits some drawbacks, distinct set of characters may have identical hash values is one of them, consequently, spurious hit occurs, leading to a mismatch.

### 2.4 Aho-Corasic Algorithm

The Aho-Corasic (AC) algorithm [12] is also popular because of its multi-pattern searching feature. This algorithm uses the idea of Finite State Automaton (FSA). The AC algorithm is proven to be linear in its search time complexity in its worst-case. The Aho-Corasic automaton is either non-deterministic finite automaton (NFA) or deterministic finite automaton (DFA). Similar to other string searching algorithms, Aho-Corasic also has two stages namely preprocessing stage and searching stage.

## 3. Proposed Algorithm

In This section we present the architecture and designing methodology of our patter recognition or identification scheme.

### 3.1 Notations

We first discuss the notations that will be used in the rest of the paper.

Table 1: Notations

| Symbols | Meaning |
|---|---|
| $T$ | Text string of length n |
| $P$ | Pattern string of length m |
| $N$ | Set of all natural numbers |
| $V_P$ | Value of the pattern |
| $V_T$ | Value of text-string |
| $V_{Ti}$ | Value of $i^{th}$ substring of length of the pattern string present in the text string. |
| $t_i$ | $i^{th}$ character of the of the text string T |
| $p_i$ | $i^{th}$ character of the of the pattern string P |
| $O$ | Set of all odd natural numbers |
| $E$ | Set of all even natural number |
| $n$ | length of the text string |
| $m$ | length of the pattern string |

### 3.2 Pattern Value Computation Formula

We introduce a novel exact string matching algorithm methodology and its many variants. Some formulae are constructed which are being utilized to develop our algorithm. We compute values of pattern strings using these formulae. The following multifarious formulae are discussed and some inherent weaknesses are also highlighted.

Let us consider following strings.

$T = t_0 t_1 t_2 \ldots t_{k-1} t_k t_{k+1} \ldots t_{n-2} t_{n-1}$

$P = p_0 p_1 p_2 \ldots p_{k-1} p_k p_{k+1} \ldots p_{n-2} p_{n-1}$

The value of the pattern is denoted by $V_p$ and is computed as follows.

$$V_p = \sum_{i=0}^{m-1} p_i * k^{i+1}, k \in N \qquad (1)$$

The above equation is not applicable for conventional exact-string searching case. Suppose, n = 1, the occurrences of mismatch are so prevalent. For instance, considering $T$ = CABACBCBABCABAC and $P$ = ABC. In reality, the pattern P must occur only at the index eight, however, the algorithm results show that the pattern is present at five more places because $A + B + C = C + A + B = B + C + A = A + C + B = B + A + C = C + B + A$.

As for the other natural values of n the applicability of this formula can also be proven to be synonymous. The prime reason behind this mismatch calamity is the 'commutative' property of 'additive' operation. This set of formulae can be useful when we look for all permutation of an input string.

$$V_p = kp_0 - \sum_{i=1}^{m-1} p_i * k^{i+1}, k \in N \qquad (2)$$

Putting n=1 in equation (4), we observe some mismatches which are illustrated in the following example. Suppose, $T$ =

$AAB$ and pattern as $P = CCB$. Then, $V_T = A - A - B = -B$ and $V_P = C - C - B = -B$. Therefore, the above phenomena is a drawback of this computation formula at n=1. There might have more such examples wherein this formula may not give an appropriate search result.

$$V_P = kp_0 + \sum_{i \in O} p_i * k^{i+1} - \sum_{i \in E} p_i * k^{i+1}, k \in N \quad (3)$$

Equation (3) exposes some inaccuracies in its search which is discussed as follows. We take, $T = ABBC$ and pattern $P = CBBA$. Then, $V_T = A + B - B + C = A + C$ and $V_P = C + B - B + A = A + C \Rightarrow V_T = V_P$, which is a flaw! There may be such examples where this formula may not determine appropriate search result.

$$V_P = kp_0 - \sum_{i \in O} p_i * k^{i+1} + \sum_{i \in E} p_i * k^{i+1}, k \in N \quad (4)$$

$$V_P = \sum_{i=0}^{m-1} p_i * (i+1)^k, k \in N \quad (5)$$

In equation (5), we detected another flaw which is as follows. We consider $T = AAADEF$ and $P = FFFDEA$.
Then, $V_T = A + 2A + 3A + 4D + 5E + 6F = 6A + 4D + 5E + 6F$ and $V_P = F + 2F + 3F + 4D + 5E + 6A = 6A + 4D + 5E + 6F$.

$$V_P = p_0 - \sum_{i=1}^{m-1} p_i * (i+1)^k, k \in N \quad (6)$$

For n = 1, equation (6) yields a flaw which is discussed by taking the example, $T = FFCFD$ and $P = DDCFD$. $V_T = F - 2F - 3C - 4F - 5D = -5F - 3C - 5D$ and $V_P = D - 2D - 3C - 4F - 5D = -5F - 3C - 5D$.

$$V_P = p_0 + \sum_{i \in O} p_i * (i+1)^k - \sum_{i \in E} p_i * (i+1)^k, k \in N \quad (7)$$

If we put $n = 1$ in equation (7), then we observed a mismatch which is as follows. We take a text string as $T = AAA$ and a pattern $P = BBB$. Then, $V_T = A + 2A - 3A = 0$ and $V_P = B + 2B - 3B = 0$.

$$V_P = p_0 - \sum_{i \in O} p_i * (i+1)^k + \sum_{i \in E} p_i * (i+1)^k, k \in N \quad (8)$$

The above formulae are also useful in designing the proposed algorithm. These are many variants of the algorithms. It is observed that from the above equations countably infinite number of pattern computation formulae can be derived. However, some inherent weaknesses in using some of these formulae are discussed. Thus, we observe that all of the above pattern-value computation formulae have some limitations at n = 1. It is also identified that with the increment of the value of n, the probability of mismatch reduces significantly. In addition, another factor that accelerates the rate of mismatch is frequent occurrence of consecutive same letters. With the increment of the length of the pattern length and text length reduces the probability of mismatch. However, the pattern computation formulae have some interesting properties and useful applications that will be discussed in the application section.

3.3 Substring Computation Method
Here, we discuss the above mentioned formulae by taking an example. It is obvious that the countably infinite number of pattern computation formulae can be derived from the above equations because for every natural number n, there exists a pattern value determination formula. Further, we present the general rule for determining substrings of length of the pattern present in the text string. We then discuss the merits and demerits of the above equations.
The generalized substring computation technique is as follows. Here we consider a particular forms of equation (6), putting n=1, we compute $V_P$.
$$V_P = p_0 - 2p_1 - 3p_2 - ... - (m+1)p_{(m-2)} - mp_{(m-1)}$$
Then, the algorithm computes all the possible substrings present in the text T of length of the pattern P. The naive method to compute all the substrings is shown below.
$$V_{T1} = t_0 - 2t_1 - 3t_2 - ... - (m+1)t_m$$
$$V_{T2} = t_1 - 2t_2 - 3t_3 - ... - (m+1)t_{m+1}$$
$$V_{T3} = t_2 - 2t_3 - 3t_4 - ... - (m+1)t_{m+1}$$

It continues to compute all the substrings, (n-m+1) number of substrings. However, using this technique resulting to high resource consumption with regard to time and memory.
Let the pattern string be $P[6] = ABCDEF$ and the text string be $T[8] = ABCDEFGH$. Now we apply equation 3 for $n=1$ in order to calculate $V_P$.
$V_P = A+B+C+D+E = 8.167+1.492+2.780+4.252+12.702 = 29.393$.
Then the algorithm finds all the substrings in the text string T of length of the pattern string P and subsequently computes all $V_{Ti}$ as follows.
$V_{T0} = A+B+C+D+E = 8.167+1.492+2.780+4.252+12.702 = 29.393$.
$V_{T1} = B+C+D+E+F = 1.492+2.780+4.252+12.702+2.228 = 23.454$.
$V_{T2} = C+D+E+F+G = 2.780+4.252+12.702+2.228+2.015 = 23.977$.
Then, the algorithm compares the pattern score with all the substring scores, recognizes the pattern where the score of the pattern string and substrings are equal, i.e., $V_P = V_T = 29.393$. Similarly, we evaluate the pattern score and all the substring scores using all the above equations for all n. However, the above mentioned substring evaluation method is not efficient because it utilizes the naive method of Brute-Force algorithm or Naive method. In order to improve the efficiency of this particular version of the algorithm [13]. We have implemented this algorithm in C that shows this technique optimizes searching time than using the naive method. A segment of the pseudocode is designed which is as follows.

Step1: [INITIALIZE] I=0;

Step2: REPEAT Step3 and Step4 while I ≤ n-1

Step3: AUX [I] ← AUX [I-1] + STR [I+m-1] -STR [I+1];

Step4: SUBSTRINGVAL [I] ← SUBSTRINGVAL [I-1] - m* STR [I_m-1] -STR [I-1] + AUX [I-1] + 3 * STR [I];

[END OF THE LOOP]

Step5: EXIT

The above segment of the pseudocode for computing each substring of size of the pattern present in the text string in an efficient way. We have taken an array to store all the computed values of the substrings and is denoted by SUBSTRINGVAL []. Further, we consider an additional or auxiliary array to store some intermediate temporary elements, denoted as AUX [].

## 4. Results

We have implemented the proposed algorithm in C programming language and python. We implemented this algorithm in a signature based Intrusion Detection System (IDS) and validated using an offline dataset. The dataset is converted into strings with the help of corresponding DNA sequences [13]. Firstly, we trained our IDS with 100,000 records with labels in the form of long strings. Then we tested the algorithm using unlabeled dataset consisting subset of strings. Following table shows the number of mismatched occurred in the experiment.

Table2: Number of Mismatches

| Number of Strings | Number of Mismatches |
|---|---|
| 10,000 | 9 |
| 20,000 | 26 |
| 30,000 | 54 |
| 40,000 | 81 |
| 50,000 | 86 |
| 60,000 | 102 |
| 70,000 | 111 |
| 80,000 | 123 |
| 90,000 | 148 |
| 100,000 | 156 |

The following graph shows how the number of mismatches occurred with the increment of number of strings.

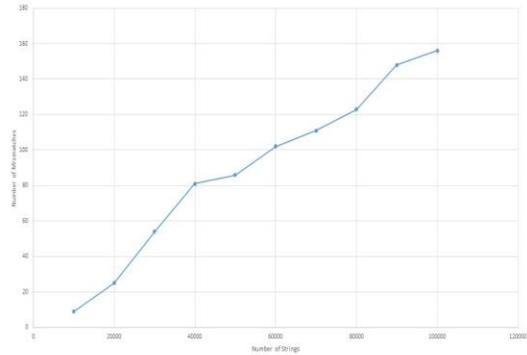

Fig. 1. Graph of mismatch occurrences

The graph shows that the number of mismatches increase with the rise of the number of strings. The above graph demonstrates that the number of false positives increase with increase of records.

## 5. Application

The proposed algorithm and its different variants have some unique properties and multifarious applications. This algorithm of ours is unique in its design method and could be employed IoT environment to detect anomaly. Whereas, the conventional pattern searching algorithms such as Boyer-Moore, Rabin-Karp and KMP algorithm have some their own merits and drawbacks as well. Furthermore, this algorithm could be applied in Bioinformatics to search DNA sequences from a large dataset efficiently. One of the variants of the algorithm can be useful in searching ubiquitous devices. The advent of internet of things (IoT) and nomenclature of IoT devices have addressed many challenges to the research community. We tried to resolve this issue taking a small scale IoT devices connected to a private network. We have discussed the naming methodology and how this algorithm helps searching all the connected devices (nodes) by their unique names.

## 6. Conclusion and Future Directions

In this paper, we have addressed a string searching algorithm based on our previous work in [13]. We also have introduced some generalized formulae to determine pattern value and from which countably many formulae can be deduced. In addition, we have shown a substring computation technique that reduces the searching complexity to O (n). The experimental result shows that the algorithm can search any length of string from a large database efficiently with minor error. Our future work involves improving our algorithm to minimize mismatches. Further, we are trying to deploy this algorithm to search edge devices by their unique names and device ID.

**Acknowledgments**

The main segment of this research work was done at Institute of Engineering & Management, Kolkata, India.